\documentclass[11pt,epsfig]{article} 
\usepackage{layout}

\usepackage[vcentermath,enableskew]{youngtab}
\usepackage{ytableau}
\usepackage{tikz}
\usepackage{enumerate}
\usepackage{float}
\usepackage{subfig}
\usepackage{color}
\usepackage{xcolor}
\usepackage{slashed}
\usepackage{multirow}
\usepackage{cite}

\let\counterwithin\relax
\usepackage{chngcntr}
\usepackage{amssymb, amsmath,mathrsfs}
\usepackage{braket, slashed, bm}

\usepackage{graphics}
\usepackage{graphicx}
\usepackage{epsf}
\usepackage{epsfig}
\usepackage{float}
\usepackage{makecell}
\usepackage{multirow}
\usepackage{color}
\usepackage{xcolor}
\usepackage{simplewick}

\usepackage[utf8]{inputenc}
\usepackage{amsmath}
\usepackage{amssymb}
\usepackage{subfig}
\usepackage[normalem]{ulem}

\usepackage{mathtools}
\usepackage{amsmath}
\usepackage{diagbox}

\usepackage{longtable}
\usepackage{ltablex,booktabs}
\usepackage{tabularray}
\newcolumntype{Y}{>{\centering\arraybackslash}X}
\newcolumntype{C}[1]{>{\centering\arraybackslash}m{#1}}
\usepackage{hhline}
\usepackage{comment}
\usepackage[bb = boondox]{mathalpha}
\usepackage{scrextend}
\usepackage{pifont}

\newcommand\undermat[2]{
	\makebox[0.5pt][l]{$\smash{\underbrace{\phantom{%
					\begin{matrix}#2\end{matrix}}}_{ \let\scriptstyle\textstyle\text{\large $#1$}}}$}#2}
\newcommand\overmat[2]{
	\makebox[-1pt][l]{$\smash{\overbrace{\phantom{%
					\begin{matrix}#2\end{matrix}}}^{ \let\scriptstyle\textstyle\text{\large $#1$}}}$}#2}    
\usepackage{tikz}
\usepackage[vcentermath]{youngtab}
\usepackage{slashed} 
\usepackage[font=small]{caption} 
\usepackage{times} 

\usepackage{bm}       
\usepackage{bbm} 

\usepackage{relsize}  

\usepackage{autobreak}  

\usepackage[colorlinks=true,
linkcolor=blue,
urlcolor=red,
citecolor=red]{hyperref}

\usepackage{makeidx} 
\usepackage{bbding} 
\usepackage{listings} 
\usepackage{ytableau}
\usepackage[utf8]{inputenc}
\usepackage[T1]{fontenc}
\usepackage{lmodern}

\usepackage[most]{tcolorbox}
\usepackage{color, colortbl}

\ytableausetup
{boxsize=1.0em}

\long\def\rpl#1!!#2!!{\textcolor{red}{#1} \textcolor{blue}{#2}}

\usepackage[top=1in, left=0.95in, bottom=1.1in, right=0.95in]{geometry}
\def\baselinestretch{1.27}
\usepackage[toc]{appendix}

\newcommand{\beq}{\begin {equation}}
\newcommand{\eeq}{\end   {equation}}
\newcommand{\bea}{\begin {eqnarray}}
\newcommand{\eea}{\end   {eqnarray}}
\newcommand{\beqa}{\begin {eqnarray}}
\newcommand{\eeqa}{\end   {eqnarray}}
\newcommand{\baa}{\begin {array}   }
\newcommand{\eaa}{\end   {array}   }
\newcommand{\bit}{\begin {itemize} }
\newcommand{\eit}{\end   {itemize} }
\newcommand{\be }{\begin {equation}}
\newcommand{\ee }{\end   {equation}}

\newcommand{\caly}{\mathcal{Y}}

\newcommand{\sEFT}{\textit{sEFT}}
\newcommand{\gEFT}{\textit{GEFT}}
\newcommand{\aEFT}{\textit{aEFT}}
\newcommand{\XEFT}{\textit{XEFT}}

\newcommand{\xcheckmark}{\checkmark\kern-1.1ex\raisebox{.7ex}{\rotatebox[origin=c]{125}{--}}}

\begin{document}

\begin{center}




{\Large \textbf  {Effective Field Theories of Axion, ALP and Dark Photon}}\\[10mm]

Huayang Song$^{a}$\footnote{huayangs@itp.ac.cn}, Hao Sun$^{a, b}$\footnote{sunhao@itp.ac.cn}, Jiang-Hao Yu$^{a, b, c, d, e}$\footnote{jhyu@itp.ac.cn}\\[10mm]

\noindent 
$^a${\em \small CAS Key Laboratory of Theoretical Physics, Institute of Theoretical Physics, Chinese Academy of Sciences,    \\ Beijing 100190, P. R. China}  \\
$^b${\em \small School of Physical Sciences, University of Chinese Academy of Sciences,   Beijing 100049, P.R. China}   \\
$^c${\em \small Center for High Energy Physics, Peking University, Beijing 100871, China} \\
$^d${\em \small School of Fundamental Physics and Mathematical Sciences, Hangzhou Institute for Advanced Study, UCAS, Hangzhou 310024, China} \\
$^e${\em \small International Centre for Theoretical Physics Asia-Pacific, Beijing/Hangzhou, China}\\[10mm]

\date{\today}   
          
\end{center}

\begin{abstract}

With the help of Young tensor technique, we enumerate the complete and independent set of effective operators up to $dim$-8 for the extension of the standard model with a Goldstone boson by further imposing the Adler's zero condition in the soft momentum limit. Such basis can be reduced to describe the axion or majoron effective Lagrangian if further (symmetry) constraints are imposed. Then reformulating dark photon as combination of Goldstone boson and transverse gauge boson, the effective operators of the Goldstone boson can be extended to effective chiral Lagrangian description of the dark photon. For the first time we obtain 0 (0), 6 (44), 1 (1), 44 (356), 32 (520) operators in Goldstone effective field theory, and 9 (49), 0 (0), 108 (676), 10 (426), 1904 (40783) operators in dark photon effective field theory at the dimension 4, 5, 6, 7, 8 for one (three) generation of fermions.
	

\end{abstract}

\newpage

\setcounter{tocdepth}{3}
\setcounter{secnumdepth}{3}

\tableofcontents

\setcounter{footnote}{0}

\def\baselinestretch{1.5}
\counterwithin{equation}{section}

\newpage

\section{Introduction}




The search for physics beyond the standard model (BSM) is one of the most pursued research avenues in high-energy physics after the discovery of the Higgs particle~\cite{ATLAS:2012yve, CMS:2012qbp}. Inspired by the experimental observations, e.g. matter-antimatter asymmetry, existence of dark matter, neutrino oscillation,
etc, and/or theoretical considerations, as e.g. supersymmetry, grand unified theory, strong CP problem, hierarchy problem, etc, models of BSM physics are constructed by augmenting the SM with new particle(s) and interaction(s) to address one or several above mentioned specific hints. With heavy degrees of freedom (DOFs) introduced, the renormalizable ultraviolet (UV) Lagrangian can be constructed to veils the new physics (NP) phenomena and real physics behind the complexity.

Given the lack of signals of BSM particles at the Large Hadron Collider (LHC), scale separation between the new physics scale and the electroweak scale is typically assumed, and thus the effective field theory (EFT) approach becomes popular in the community recent years. EFT framework describe physics at a specific energy scale by parameterizing the impact of physics at other scales into the free parameters (Wilson coefficients) of the theory. Therefore they are constructed by identifying the only relevant fields that physics phenomena can be characterised at the interesting scale. Typically EFTs contain an infinite tower of higher dimensional operators built from the light DOFs invariant under the required symmetries and organized under some power counting rule, however one should only focus on a few lower dimensional operators relevant to the phenomenology that are considered, which significantly simplifies the scenario and highlight the underlying physics. In specific, the physics below and at the electroweak (EW) scale can be described by the standard model effective field theory (SMEFT), which is only composed of all the SM fields and the operator bases have been presented in literature~\cite{Weinberg:1979sa, Buchmuller:1985jz, Grzadkowski:2010es, Lehman:2014jma, Henning:2015alf, Liao:2016hru, Li:2020gnx, Murphy:2020rsh, Li:2020xlh, Liao:2020jmn}. 

However, the lack of BSM signal at the LHC does not rule another possibility out, light new particles below GeV, or feebly interacting particles (FIPs) below or around the EW scale, such as axion or axion-like particles (ALPs)~\cite{Peccei:1977hh, Peccei:1977ur, Weinberg:1977ma, Wilczek:1977pj}, light scalar particles~\cite{Espinosa:1993bs, OConnell:2006rsp, Barger:2006sk, Ahriche:2007jp, Profumo:2007wc, Barger:2007im, Barger:2008jx, Gonderinger:2009jp, Espinosa:2011ax, Pruna:2013bma, Chen:2014ask, Gorbahn:2015gxa, Dawson:2015haa, Costa:2015llh, Robens:2016xkb, Patt:2006fw}, not-so-heavy neutral leptons (HNLs)~\cite{Minkowski:1977sc, Yanagida:1979as, Gell-Mann:1979vob, Mohapatra:1979ia} and dark photons~\cite{Okun:1982xi, Galison:1983pa, Holdom:1985ag}. Such kind of particles can be directly produced with enough statistics at the high luminosity run of the LHC and observed by the traditional detectors ATLAS, CMS and LHCb via missing transverse energy or displaced vertex signature~\cite{Gershtein:2017tsv, Liu:2018wte, Lee:2018pag, Alimena:2019zri, Liu:2019ayx, Liu:2020vur, Gershtein:2020mwi, Chiu:2021sgs, Fischer:2021sqw, Bose:2022obr}, or by CODEX-b~\cite{Gligorov:2017nwh, Dey:2019vyo, Aielli:2019ivi, Aielli:2022awh}, FASER~\cite{Feng:2017uoz, FASER:2018ceo, FASER:2018bac, FASER:2022hcn, FASER:2021ljd,FASER:2021cpr}, MATHUSLA~\cite{Chou:2016lxi, Curtin:2017izq, Evans:2017lvd, Curtin:2018mvb, Curtin:2018ees, MATHUSLA:2018bqv,  MATHUSLA:2019qpy, Alimena:2019zri, Alidra:2020thg, MATHUSLA:2020uve, MATHUSLA:2022sze, Bose:2022obr}, et al, if they are long-lived due to the light masses or their suppressed interactions. To study the phenomena of these particles, it is more convenient to provide a model-independent way to describe the light new particle or FIP extensions of the SM, which extend the SMEFT to EFTs with light particles. These particles and their EFTs could also be produced via meson decays and tested in fixed target experiments such as NA62/64~\cite{Lanfranchi:2017wzl, NA62:2017rwk, NA64:2016oww, Gninenko:2019qiv, Banerjee:2019pds, NA64:2020qwq, Gninenko:2300189, Gninenko:2640930}, SHiP~\cite{Bonivento:2013jag, Alekhin:2015byh, SHiP:2015vad}, or Sea/Spin/Darkquest~\cite{SeaQuest:2017kjt, Liu:2017ryd, Berlin:2018pwi, Batell:2020vqn, Blinov:2021say, Apyan:2022tsd, Tsai:2019buq} at the high luminosity frontier.

Generally the particles carrying the SM quantum numbers are forced to interact with the SM particles via SM gauge interactions, which are not feebly enough, and consequently the SM gauge singlets are easier to evade the severe experimental constraints below the TeV scale. Spontaneous symmetry breaking generates massless scalar particles, which are known as Nambu-Goldstone bosons. Though explicit breaking terms can be introduced to lift the masses of Goldstones, they typically remain relatively light. Such SM singlet particles (pseudo-Nambu-Goldstone bosons, or pNGBs), therefore, are natural extensions of the SMEFT which need to be studied. The Goldstones can further work as the longitudinal modes of vector bosons if some symmetries are gauged. Inspired by the Goldstone equivalence theorem~\cite{Chanowitz:1985hj, Vayonakis:1976vz, Cornwall:1974km, Lee:1977eg, Gounaris:1986cr, Yao:1988aj, Veltman:1989ud, Bagger:1989fc} and the development of Higgs effective field theory (HEFT)~\cite{Appelquist:1980vg, Longhitano:1980iz, Longhitano:1980tm, Feruglio:1992wf, Herrero:1993nc, Herrero:1994iu, Grinstein:2007iv, Buchalla:2012qq, Buchalla:2013rka, Buchalla:2013eza, Gavela:2014vra, Pich:2015kwa, Pich:2016lew, Krause:2018cwe, Alonso:2012px, Brivio:2013pma, Brivio:2016fzo, Pich:2018ltt, Merlo:2016prs, Sun:2022ssa, Sun:2022snw}, we realize that the EFT of Goldstone singlet can also describe the interactions of dark photon by certain modification. Hence in this work we focus on the Goldstone singlet extension, its variations and dark photon extension. The most general EFT operators of the singlet SMEFT extensions with arbitrary spins are given in a companion paper~\cite{Song:2023jqm}.

ALPs are a special class of pseudo-Goldstone bosons, and ALP-SMEFT effective couplings are well-studied previously~\cite{Georgi:1986df, Brivio:2017ije, Bauer:2017ris, Bauer:2018uxu, Chala:2020wvs, Bauer:2020jbp, Bonilla:2021ufe, Brivio:2021fog, Galda:2021hbr, Arina:2021nqi, Ghosh:2023tyz}. However, those studies concentrate on the lowest level interactions (mostly $dim$-5 operators, some also include the only one $dim$-6 operator), how to identify the non-redundant operators, operator transformation between different bases, one-loop corrections to these couplings, and the corresponding collider signatures. Fully general SMEFT extensions including a Goldstone singlet (\gEFT) beyond the leading level are not known. One should note that the lowest dimensional operators ($dim$-5 or -6) can be naturally forbidden by some non-trivial symmetries of these particles (e.g. as will be shown later, the leading order operators involving a majoron who carries $B-L$ charge $2$ show up at dimension 8). Further the higher dimensional operators may lead to detectable level effects at the high luminosity frontier. For instance, the exotic Higgs decay $h\rightarrow Za$ is mediated by a dimension 7 operator $(\partial^\mu a)(H^\dagger\partial_\mu H)H^\dagger H$ mediates at the leading order~\cite{Bauer:2016zfj, Bauer:2016ydr, Bauer:2017ris}. Therefore for the first time, we construct the complete and independent operators involving SM fields and an extra SM singlet of Goldstone up to dimension eight using Young tensor technique developed in Ref.~\cite{Li:2020gnx, Li:2020xlh, Li:2020zfq, Li:2022tec} with certain improvements dealing with Goldstone~\cite{Sun:2022ssa, Sun:2022snw}. Appending to the \gEFT~with an extra $U(1)$ gauge boson, we can also build higher dimensional operators for a dark photon. Such kind of EFT for a massive vector singlet (\XEFT) has only been discussed up to dimension $\leq 6$ in the framework of dark matter in Ref.~\cite{Duch:2014xda, Criado:2021trs} with a $\mathbb{Z}_2$ symmetry to stabilize DM particle and in Ref.~\cite{Barducci:2021egn, Aebischer:2022wnl} without imposing any symmetries. Ref.~\cite{Kribs:2022gri} investigates the effective field theory of a vector field that has a Stueckelberg mass at the $dim$-4 level and examines the high energy behavior of the scattering amplitudes at the one-loop order. Based on our construction, arbitrary higher dimensional operators for \XEFT~can be obtained without redundancy. 

The paper is organized as follows. In Sec.~\ref{sec:opsbbs&YTT} we briefly review the building blocks and the Young tensor method to construct the complete and independent effective operator, and also discuss how to obtain operators involving Goldstone by imposing Adler's Zero condition. Based on these, we list the complete operators of \gEFT~up to $dim$-8 in Sec.~\ref{sec:operator4goldstone}. In Sec.~\ref{sec:operator4photon} we firstly review the Stueckelberg and Higgs mechanism in dark photon model, and then describe how a massive vector boson EFT can be built via Stueckelberg mechanism and Goldstone equivalent theorem by introducing both a $U(1)$ gauge field and a Goldstone boson and present our new power-counting scheme. The complete operators up to dimension eight of single singlet extension of the SMEFT with a vector (spin-1) particle are listed at the end. We reach our conclusion in Sec.~\ref{sec:conclu}.

\section{Effective Lagrangian of Singlet Extensions}
\label{sec:opsbbs&YTT}
In this section, we lay out the main ingredients of the singlet extensions of the SMEFT: the particle content (build blocks), and their global and local symmetries. We briefly summarize the Young Tensor technique of effective operator construction in Sec.~\ref{sec:YTT}, which was implemented by Mathematica package~\cite{Li:2022tec}. The shift symmetry of the Goldstone is taken care of by imposing the Adler zero conditions on the amplitudes, as explained in Sec.~\ref{sec:Adler}.

\subsection{Classification of Singlet Particles}
In the construction of effective operators, we adopt the fields of the irreducible representations of the Lorentz group as the DOFs. Up to spin-$1$, these fundamental building blocks and their representations are 
\begin{align}
    h=0: & \quad\phi\in (0,0)\notag\\
    h=-\frac{1}{2}: &\quad \psi\in(\frac{1}{2},0)\notag\\
    h=\frac{1}{2}: &\quad \psi^\dagger\in(0,\frac{1}{2}) \notag \\
    h=-1: &\quad X_L\in(1,0) \notag \\
    h=1: &\quad X_R\in(0,1) \,,
\end{align}
in terms of helicities $h$, where $\psi$ is left-handed fermions, $\psi^\dagger$ is right-handed fermions. $X_{L/R}$ is a left/right-handed field-strength tensor, which is related to the normal tensors by
\begin{equation}
    X_{L/R}^{\mu\nu} = \frac{1}{2}(X^{\mu\nu}\mp i \tilde{X}^{\mu\nu})\,,
\end{equation}
where $\tilde{X}^{\mu\nu}=\epsilon^{\mu\nu\rho\lambda}X_{\rho\lambda}$.

In the on-shell cases, the fermions correspond to Weyl spinors,
\begin{equation}
\label{eq:fermion}
    \psi \rightarrow \lambda_\alpha\,, \quad\psi^\dagger \rightarrow \tilde{\lambda}^{\dot{\beta}}\,,
\end{equation}
where the undotted Greek characters are used as the indices of the $SU(2)_l$ and the dotted ones are used as the indices of the $SU(2)_r$.\footnote{We use the cover group of the Lorentz group $SU(2)_l\times SU(2)_r\sim SO(3,1)$, of which the irreducible representations are noted by $(j_l,j_r)$, where $j_l,j_r$ are the Casimir's eigenvalues of $SU(2)_l\,, SU(2)_r$ respectively.} According to the tensor product rule of $SU(2)$ group, 
\begin{equation}
    \mathbf{\frac{1}{2}}\otimes\mathbf{\frac{1}{2}} = \mathbf{1}\oplus \mathbf{0}\,,
\end{equation}
the irreducible representation $(1,0)/(0,1)$ of the Lorentz group is the symmetric part of the tensor product of two $(1/2,0)/(0,1/2)$ representations. Thus the left- and right-handed field-strength tensors can be parameterized by the Weyl spinors as their symmetric products
\begin{equation}
\label{eq:vector}
    X_{L} \rightarrow \lambda_\alpha \lambda_\beta\,,\quad X_{R} \rightarrow \tilde{\lambda}^{\dot{\alpha}}\tilde{\lambda}^{\dot{\beta}}\,.
\end{equation}
Besides, the derivative $D$, or the on-shell momentum $p$, is of the representation $(1/2,1/2)$, thus corresponds to the spinor form 
\begin{equation}
\label{eq:derivative}
    D \rightarrow \lambda_\alpha \tilde{\lambda}^{\dot{\beta}}\,.
\end{equation}
In particular, the Goldstone boson due to symmetry breaking described by non-linear symmetry realization is special since it satisfies Adler zero condition~\cite{Adler:1964um,Adler:1965ga}, which means there is at least one derivative applying on it. Assuming $\phi$'s are Golstone bosons, they are collect in the unitary matrix $\mathbf{U}(x)$~\cite{Weinberg:1978kz,Gasser:1983yg,Gasser:1984gg}, and participate in the dynamics in the form of 
\begin{equation}
    u_\mu = i\mathbf{U} D_\mu\mathbf{U}^\dagger\,,
\end{equation}
thus it is of the same representation of derivative,
\begin{equation}
\label{eq:gbs}
    u_\mu \rightarrow \lambda_\alpha\tilde{\lambda}^{\dot{\beta}} \in (\frac{1}{2},\frac{1}{2})\,.
\end{equation}

According to the discussion above, we can interpret all the fields in the SM together with the various singlet extensions to spinor forms. Restoring the SM gauge symmetry $SU(3)_C\times SU(2)_W\times U(1)_Y$ and specific added abelian symmetry $U(1)_A$, we list all the building blocks in Tab.~\ref{tab:buildingblocks}, where the charges $q_S\,,q_u\,,q_M\,,q_D$ of the singlet particles under the added $U(1)_A$ symmetry are assumed.

\begin{table}[]
    \centering
 \begin{tabular}{c|c|c|c|c|c||c}
\hline
Sector & Building block & Lorentz group & $SU(3)_C$ & $SU(2)_W$ & $U(1)_Y$ & $U(1)_A$\\
\hline
\multirow{9}{*}{SM} & $G_{L/R}$ & $(1,0)/(0,1)$ & $\mathbf{8}$ & $\mathbf{1}$ & 0 & 0 \\ 
\cline{2-7}
& $W_{L/R}$ & $(1,0)/(0,1)$ & $\mathbf{1}$ & $\mathbf{3}$ & 0 & 0 \\
\cline{2-7}
& $B_{L/R}$ & $(1,0)/(0,1)$ & $\mathbf{1}$ & $\mathbf{1}$ & 0 & 0 \\
\cline{2-7}
& $L/L^\dagger$ & $(\frac{1}{2},0)/(0,\frac{1}{2})$ & $\mathbf{1}$ & $\mathbf{2}$ & $\mp \frac{1}{2}$ & 0 \\
\cline{2-7}
& $e_c/e_c^\dagger$ & $(\frac{1}{2},0)/(0,\frac{1}{2})$ & $\mathbf{1}$ & $\mathbf{1}$ & $\pm 1$ & 0 \\
\cline{2-7}
& $Q/Q^\dagger$ & $(\frac{1}{2},0)/(0,\frac{1}{2})$ & $\mathbf{3}/\overline{\mathbf{3}}$ & $\mathbf{2}$ & $\pm \frac{1}{6}$ & 0 \\
\cline{2-7}
& $u_c/u_c^\dagger$ & $(\frac{1}{2},0)/(0,\frac{1}{2})$ & $\overline{\mathbf{3}}/\mathbf{3}$ & $\mathbf{1}$ & $\mp \frac{2}{3}$ & 0 \\
\cline{2-7}
& $d_c/d_c^\dagger$ & $(\frac{1}{2},0)/(0,\frac{1}{2})$ & $\overline{\mathbf{3}}/\mathbf{3}$ & $\mathbf{1}$ & $\pm \frac{1}{3}$ & 0 \\
\cline{2-7}
& $H/H^\dagger$ & $(0,0)$ & $\mathbf{1}$ & $\mathbf{2}$ & $\pm \frac{1}{2}$ & 0 \\
\hline
\hline
Real scalar & $s$ & $(0,0)$ & $\mathbf{1}$ & $\mathbf{1}$ & 0 & 0 \\
\hline
\hline
Complex Scalar & $S/S^\dagger$ & $(0,0)$ & $\mathbf{1}$ & $\mathbf{1}$ & 0 & $\pm q_S$ \\
\hline
\hline
Goldstone & $u_\mu \sim D_\mu s$ & $(\frac{1}{2},\frac{1}{2})$ & $\mathbf{1}$ & $\mathbf{1}$ & 0 & 0 \\
\hline
\hline
Majorana fermion & $\chi/\chi^\dagger$ & $(\frac{1}{2},0)/(0,\frac{1}{2})$ & $\mathbf{1}$ & $\mathbf{1}$ & 0 & $\pm q_M$ \\
\hline
\hline
\multirow{2}{*}{Dirac fermion} & $\chi_L/\chi^\dagger_L$ & $(\frac{1}{2},0)/(0,\frac{1}{2})$ & $\mathbf{1}$ & $\mathbf{1}$ & 0 & $\pm q_D$ \\
\cline{2-7}
& $\chi_{Rc}/\chi_{Rc}^\dagger$ & $(\frac{1}{2},0)/(0,\frac{1}{2})$ & $\mathbf{1}$ & $\mathbf{1}$ & 0 & $\mp q_D$ \\
\hline
\hline
dark photon & $X_L/X_R$ & $(1,0)/(0,1)$ & $\mathbf{1}$ & $\mathbf{1}$ & 0 & 0 \\
\hline
\end{tabular}
    \caption{The fields in this table are classified into several parts, the fields of the SM, the scalar extension, the Goldstone extension, the Majorana fermion extension, the Dirac fermion extension, and the dark photon extension, where $q_S\,,q_u\,,q_M\,,q_D$ are the charges of the complex particles under the added $U(1)_A$ symmetry. }
    \label{tab:buildingblocks}
\end{table}

\subsection{Young Tensor Technique}\label{sec:YTT}
In terms of the spinor forms of the fields, the independent and complete effective operators can be constructed systematically via the so-called Young tensor method~\cite{Henning:2019enq, Henning:2019mcv, Li:2020gnx, Li:2020xlh, Li:2022tec}. In this section, we shall review the basic ingredients of this systematic method, and more details and examples can be found in Ref.~\cite{Li:2020gnx, Li:2020xlh, Li:2022tec}. Since all the effective operators are invariant under both the Lorentz group and gauge group, we will discuss the two sectors in sequence, in both of which the Young diagram techniques play the central role.

In the Lorentz sector, the invariant means that all the spinors are contracted by the anti-symmetric tensor $\epsilon/\tilde{\epsilon}$. Defining
\begin{equation}
\epsilon^{\alpha\beta}\lambda_\beta\xi_\alpha = \lambda^\alpha\xi_\alpha = \langle\lambda\xi\rangle\,,\quad \epsilon_{\dot{\alpha}\dot{\beta}}\tilde{\lambda}^{\dot{\beta}}\tilde{\xi}^{\dot{\alpha}} = \tilde{\lambda}_{\dot{\alpha}}\tilde{\xi}^{\dot{\alpha}} = [\tilde{\lambda}\tilde{\xi}]\,,
\end{equation}
which satisfy anti-symmetric relation\footnote{These spinors are not Grassmannian.}
\begin{equation}
\langle\lambda\xi\rangle = -\langle\xi\lambda\rangle\,,\quad [\lambda\xi]=-[\xi\lambda]\,,
\end{equation}
thus a general Lorentz invariant is an amplitude composed of these brackets
\begin{equation}
\mathcal{D} = \prod^n\langle\,.\,\rangle \prod^{\tilde{n}} [\,.\,]\,,
\end{equation}
where the $n\,,\tilde{n}$ is the number of the two kinds of brackets respectively. For a given fields contents with some derivatives, which is also called types, although the $n\,,\tilde{n}$ are determined according to their corresponding spinor forms presented in the previous section, there are still many relations, containing momentum conservation, Schouten identities, make it difficult to construct the independent and complete Lorentz structures. Nevertheless, assuming there are $N$ fields and $n_D$ derivatives in the type, with helicities $h_1,h_2\,,\dots,h_N$. It has been argued~\cite{Henning:2019enq,Henning:2019mcv} all the independent structures are captured by a single Young diagram of the shape
\begin{equation}
    \begin{tikzpicture}
\filldraw [draw = black, fill = cyan] (10pt,10pt) rectangle (22pt,22pt);
\filldraw [draw = black, fill = cyan] (10pt,22pt) rectangle (22pt,34pt);
\draw [densely dotted] (24pt,22pt) -- (32pt,22pt);
\filldraw [draw = black, fill = cyan] (10pt,46pt) rectangle (22pt,58pt);
\filldraw [draw = black, fill = cyan] (10pt,58pt) rectangle (22pt,70pt);
\draw [densely dotted] (24pt,58pt) -- (32pt,58pt);
\draw [densely dotted] (16pt,36pt) -- (16pt,46pt);
\filldraw [draw = black, fill = cyan] (34pt,10pt) rectangle (46pt,22pt);
\filldraw [draw = black, fill = cyan] (34pt,22pt) rectangle (46pt,34pt);
\filldraw [draw = black, fill = cyan] (34pt,46pt) rectangle (46pt,58pt);
\filldraw [draw = black, fill = cyan] (34pt,58pt) rectangle (46pt,70pt);
\draw [densely dotted] (40pt,36pt) -- (40pt,46pt);
\draw [|<-] (0pt,10pt)--(0pt,34pt);
\draw [|<-] (0pt,70pt)--(0pt,46pt);
\node (n2) at (-5pt,40pt) {\small $N-2$};

\draw [|<-] (10pt,80pt)--(22pt,80pt);
\draw [|<-] (46pt,80pt)--(34pt,80pt);
\node (nt) at (28pt,80pt) {\small $\tilde{n}$};

\draw (46pt,58pt) rectangle (58pt,70pt);
\draw (46pt,46pt) rectangle (58pt,58pt);
\draw [densely dotted] (59pt,58pt) -- (67pt,58pt);
\draw (68pt,58pt) rectangle (80pt,70pt);
\draw (68pt,46pt) rectangle (80pt,58pt);
\draw [|<-] (46pt,36pt)--(58pt,36pt);
\draw [|<-] (80pt,36pt)--(68pt,36pt);
\node (nt) at (63pt,36pt) {\small $n$};
    \end{tikzpicture}\,,
\end{equation}
called primary Young diagram, 
and a specific basis can be obtained by filling numbers in that to obtain the semi-standard Young tableaux (SSYT), which requires all the numbers in the same row to increase weakly, while all the numbers in the same column increase strictly. The numbers $i$ should be filled ranging from 1 to the field number $N$, and the repetition of each of them is determined by  
\begin{equation}
\#i = \frac{1}{2}n_D+\sum_{h_i>0}|h_i|-2h_i,\quad i=1,2,\dots N\,.
\end{equation}
Given SSYTs, the independent Lorentz structures can be obtained by the interpretation column by column, 
\begin{equation}
\begin{ytableau}
*(cyan) i  \\
*(cyan) i  \\
\none[\vdots] \\
*(cyan) k \\
\end{ytableau} \rightarrow \epsilon^{ij\dots klm}[lm]\,,\quad
\begin{ytableau}
i  \\
j  \\
\end{ytableau} \rightarrow \langle ij \rangle\,.
\end{equation}

The utilization of the primary Young diagram means that the independent Lorentz structures can be obtained immediately once the type is known, which simplifies the effective operator construction, and similar situation appears in the gauge sector. 

The gauge structures of the effective operators are invariants as well, which correspond to the Young diagram
\begin{equation}
\label{eq:singletYD}
\begin{tikzpicture}
\draw (20pt,20pt) rectangle (34pt,34pt);
\draw (34pt,20pt) rectangle (48pt,34pt);
\draw [loosely dotted] (50pt,27pt)--(60pt,27pt);
\draw (62pt,20pt) rectangle (76pt,34pt);
\draw [loosely dotted] (27pt,36pt)--(27pt,46pt);
\draw [loosely dotted] (69pt,36pt)--(69pt,46pt);
\draw (20pt,48pt) rectangle (34pt,62pt);
\draw (34pt,48pt) rectangle (48pt,62pt);
\draw [loosely dotted] (50pt,62pt)--(60pt,62pt);
\draw (62pt,48pt) rectangle (76pt,62pt);
\draw (20pt,62pt) rectangle (34pt,76pt);
\draw (34pt,62pt) rectangle (48pt,76pt);
\draw (62pt,62pt) rectangle (76pt,76pt);
\draw [|<-] (13pt,20pt)--(13pt,41pt);
\draw [|<-] (13pt,76pt)--(13pt,55pt);
\node (N) at (13pt,48pt) {{\small$N$}};
\end{tikzpicture}
\end{equation}
for $SU(N)$ group, called singlet Young diagram.
Every irreducible representation of $SU(N)$ corresponds to a Young tableau, the tensor product of them can be performed by the outer product, which respects the Littlewood-Richardson rule. Thus the independent $SU(N)$ invariants can be obtained by constructing the Young tableau of the shape of a singlet Young diagram from the Young diagrams of the fields in terms of outer product, and the SSYTs obtained in such a way are interpreted to group tensors by transferring every column to $N$-rank anti-symmetric tensor,
\begin{equation}
\begin{ytableau}
i  \\
j  \\
\none[\vdots] \\
k \\
\end{ytableau} \rightarrow \epsilon^{ij\dots k}\,.
\end{equation}
 
Given the independent structures of these two sectors, The tensor product of them forms the complete and independent operator basis if there are no repeating fields, called the m-basis. In the case of repeating fields, the flavor structures of the operators can be complicated, and some operators of specific flavor symmetry may not be independent, thus it is necessary to combine all the resultant operators to form the ones with specific flavor symmetry and eliminate the possible redundancies. The efficient way to do this is to construct the idempotent of the flavor indices, $\mathcal{Y}^{[\lambda]}$, 
where $[\lambda]$ is the corresponding Young diagram, representing specific flavor symmetry, and the rank of the idempotent implies the number of independent operators. After the applications of the idempotents on the m-basis, the truly independent operators form the so-called f-basis, organized by flavor symmetry.

In summary, the Young tensor method is a systematic method that not only constructs effective operators up to any dimension generally but also can determine their complete flavor symmetry. It has been applied to SMEFT~\cite{Li:2020gnx, Li:2020xlh}, LEFT~\cite{Li:2020tsi}, and HEFT~\cite{Sun:2022ssa, Sun:2022snw}. In this paper, this method will be used to construct the effective operators of the SMEFT with various extensions. In addition, we adopt the flavor indices $\{p,r,s,t,\dots\}$ in this paper.
In the case of repeating fields, the permutation symmetry of their flavor indices is represented by the idempotent of the form
\begin{equation}
\mathcal{Y}[{\tiny\ytableaushort{pr}}]\,,\quad \caly[{\tiny\ytableaushort{s,t}}]\,,\quad \caly[{\tiny\ytableaushort{pr},\ytableaushort{s,t}}]
\end{equation}
and so on.

\subsection{Goldstone Bosons}\label{sec:Adler}
As discussed previously, the Goldstone bosons are special since they satisfy Adler's zero condition. Although we have argued that the DOF representing the Goldstone bosons is $u_\mu$, as shown in Eq.~(\ref{eq:gbs}), it needs extra management in the Young tensor method. From the operator-amplitude correspondence, an operator involving the Goldstones corresponds to an on-shell amplitude satisfying the Adler's zero condition in the soft momentum limit~\cite{Adler:1964um, Adler:1965ga, Low:2014nga, Low:2014oga, Cheung:2014dqa, Cheung:2015ota, Low:2019ynd, Dai:2020cpk, Low:2022iim, Sun:2022ssa, Sun:2022snw},
\begin{align}
    \mathcal{M}(p_\pi)\rightarrow p_\pi\quad\text{for}\quad p_\pi\rightarrow 0\,,
\end{align}
where we use $p_\pi$ to represent the momentum of a general external Goldstone. This means for a given type involving the Goldstone, the Lorentz structures obtained from the primary Young diagram should satisfy the same relation. In terms of the Young tensor method, we need to restrict the Lorentz structures as follows.

Suppose the Lorentz basis is of D-dimension of the type, $\{\mathcal{B}^i|i=1,2,\dots,D\}$, they do not satisfy Adler's zero condition generally. Nevertheless, we assume their certain combinations $\sum_ic_i\mathcal{B}^i$ do satisfy the condition, which means that
\begin{equation}
    \lim_{p_\pi\rightarrow 0}\sum_{i=1}^Dc_i\mathcal{B}^i=\sum_{i=1}^Dc_i\lim_{p_\pi\rightarrow 0}\mathcal{B}^i=0\,,
\end{equation}
where $c_i$'s are undetermined coefficients, and the structures $\lim_{p_\pi\rightarrow 0}\mathcal{B}^i$ in the summation are not independent generally and can be expanded by the original basis $\{\mathcal{B}^i|i=1,2,\dots,D\}$,
\begin{equation}
    \lim_{p_\pi\rightarrow 0}\mathcal{B}^i= \sum_{j=1}^DK^{i}_j\mathcal{B}^j\,.
\end{equation}
Substituting it in the original relation we can get
\begin{equation}
    \sum_{i=1}^Dc_i(\sum_{j=1}^DK^i_j\mathcal B^j) = \sum_{j=1}^D(\sum_{i=1}^Dc_iK^i_j)\mathcal{B}^j=0\,.
\end{equation}
Since the basis $\mathcal{B}^j$'s are independent, the relation is equivalent to 
\begin{equation}
\label{eq:linear}
    \sum_{i=1}^Dc_iK^i_j=0\,,
\end{equation}
which is a system of linear equations and its solution space constrains the combination coefficients $c_i$. In particular, the dimension of the solution is the number of independent structures satisfying Adler's zero condition. We refer the readers to the original papers Ref.~\cite{Sun:2022ssa, Low:2022iim, Sun:2022snw} where this construction method was developed for detailed discussions and some explicit examples.

In summary, Adler's zero condition makes the Lorentz basis smaller, the structures of which are generally the combinations of the original basis, and the combination coefficients are determined by the system of linear equations such as Eq.~(\ref{eq:linear}). The original Lorentz basis can be further converted into a new Lorentz basis which contains all the structures satisfying the Adler's zero condition and other structures~\footnote{We present operator basis for a real scalar built from this new Lorentz basis, in which the Axion basis can be embedded.}.

\section{Goldstone EFT Operators: Axion, ALP and Majoron\label{sec:operator4goldstone}}
The Goldstone extension of the SMEFT can be identified as a specific class of scalar singlet extensions (\sEFT) in which the scalar is not a fundamental particle but a Goldstone boson emerging from spontaneous breaking of a global symmetry~\cite{Nambu:1960tm, Goldstone:1961eq, Goldstone:1962es}. Due to the shift symmetry of the Goldstone, there are no other renormalizable terms in the Lagrangian besides the kinetic term~\cite{Coleman:1969sm, Callan:1969sn}. And as mentioned in Sec.~\ref{sec:Adler}, we can obtain the \gEFT~operator basis via reducing the \sEFT~basis by imposing Adler's zero condition.

\label{hyref:axion}The most well-known particle in this class is axion (QCD axion), the Goldstone arising from the spontaneously broken global Peccei-Quinn symmetry $U(1)_\text{PQ}$, which was introduced to solve the ``strong CP problem''~\cite{Peccei:1977hh, Peccei:1977ur, Weinberg:1977ma, Wilczek:1977pj}. Such kind of particles extend well beyond the QCD axion, though. They appear in a plethora of beyond the SM (BSM) constructions, e.g. theories with extra dimensions and string theory models~\cite{Svrcek:2006yi, Arvanitaki:2009fg, Cicoli:2012sz}, which are often named as axion-like particles (ALPs). To the leading order, the complete and non-redundant Lagrangian (\aEFT) is~\cite{Georgi:1986df}
\begin{align}
    \mathcal{L}\supset&\frac{1}{2}\partial_\mu a \partial^\mu a+c_{\widetilde{B}}\frac{a}{f_a}B_{\mu\nu}\widetilde{B}^{\mu\nu}+c_{\widetilde{W}}\frac{a}{f_a}W_{\mu\nu}^a\widetilde{W}^{a\mu\nu}+c_{\widetilde{G}}\frac{a}{f_a}G_{\mu\nu}^\lambda\widetilde{G}^{\lambda\mu\nu} \nonumber \\
    &+c_u\frac{\partial_\mu a}{f_a}\left(\overline{u}_R\gamma^\mu u_R\right)+c_d\frac{\partial_\mu a}{f_a}\left(\overline{d}_R\gamma^\mu d_R\right)+c_e\frac{\partial_\mu a}{f_a}\left(\overline{e}_R\gamma^\mu e_R\right) \nonumber \\
    &+c_Q\frac{\partial_\mu a}{f_a}\left(\overline{Q}_L\gamma^\mu Q_L\right)_{i, j\neq 1, 1}+c_L\frac{\partial_\mu a}{f_a}\left(\overline{L}_L\gamma^\mu L_L\right)_{i\neq j}\,,
    \label{eq:Laxionshift}
\end{align}
where $f_a$ is the decay constant and $\widetilde{B}, \widetilde{W}, \widetilde{G}$ are the dual field tensors of $B, W, G$ respectively. In reality, QCD axion or ALPs should be pseudo-Goldstone Bosons to evade the strong constraints on massless states, therefore a mass term and even some other shift symmetry breaking terms, which however are usually ignored in analysis to highlight the Goldstone nature of the axion, can be added. Here we will also focus on the Goldstone nature of the axion while a more general EFT involving these breaking term can be treated as an EFT for a real scalar which can be found in our companion paper~\cite{Song:2023}. Axion not only provides a solution to the strong CP problem but also can serve as a dark matter candidate producing via either the misalignment mechanism or topological defects~\cite{Marsh:2015xka}.

For the axion $U(1)$ gauge field interaction, we can integrate by parts and at the same time use Bianchi identity to obtain
\begin{align}
    \frac{a}{f_a}F_{\mu\nu}\widetilde{F}^{\mu\nu}\longrightarrow -\frac{\partial_\mu a}{f_a}2A_\nu\widetilde{F}^{\mu\nu}\,,
\end{align}
which explicitly shows that axion field $a$ has a shift symmetry. In a more general way, the axion interactions to gauge bosons can be written as $\partial_\mu a K_F^\mu$ by integration by parts
\begin{align}
    \frac{a}{f_a}F_{\mu\nu}\widetilde{F}^{\mu\nu}\longrightarrow -\frac{\partial_\mu a}{f_a}K_F^\mu=-\frac{\partial_\mu a}{f_a} 2\epsilon^{\mu\nu\rho\sigma}\left(A_\nu\partial_\rho A_\sigma-\frac{2i}{3}A_\nu A_\rho A_\sigma\right)\,,
\end{align}
where $K_F^\mu$ is the Chern-Simons current associated to the $F=\{B, W, G\}$ gauge boson~\footnote{Note that there is only the first term for $U(1)_Y$ gauge field $B$.}. In this form, the shift symmetry of axion is made explicitly, however, we know that the continuous shift symmetry is broken by instanton effects to the discrete subgroup.

The equations of motion for chiral fermions read
\begin{gather}
    i\slashed{D}Q_{Lk}^\alpha=Y_d^{\alpha\beta}H_k d_R^\alpha+Y_u^{\alpha\beta}\widetilde{H}_k u_R^\beta, \qquad i\slashed{D}L_{Lk}^\alpha=Y_e^{\alpha\beta}H_k e_R^\alpha \\
    i\slashed{D}u_{Rk}^\alpha=Y_u^{\beta\alpha*} \widetilde{H}^\dagger_k Q_{Lk}^\beta, \qquad i\slashed{D}d_R^\alpha=Y_d^{\beta\alpha*}H^\dagger_k Q_{Lk}^\beta, \qquad i\slashed{D}e_R^\alpha=Y_e^{\beta\alpha*}H^\dagger_k L_{Lk}^\beta\,,
\end{gather}
with which one could rewrite the interactions between axion and fermions into non-derivative form
\begin{align}
    \mathcal{L}\supset&c_{\widetilde{B}}\frac{a}{f_a}B_{\mu\nu}\widetilde{B}^{\mu\nu}+c_{\widetilde{W}}\frac{a}{f_a}W_{\mu\nu}^a\widetilde{W}^{a\mu\nu}+c_{\widetilde{G}}\frac{a}{f_a}G_{\mu\nu}^\lambda\widetilde{G}^{\lambda\mu\nu} \nonumber \\
    &-\frac{a}{f_a}\left(c_{QHu}\overline{Q}_L\widetilde{H}u_R+c_{QHd}\overline{Q}_L Hd_R+c_{LHe}\overline{L}_L He_R+\text{h.c.}\right)\,, \label{eq:Laxionyukawa}
\end{align}
where the three couplings $c_{QHu, QHd, LHe}$ are fixed by the SM Yukawa and shift-invariant couplings $c_{Q, L, u, d, e}$
\begin{align}
    c_{QHu}=i\left(Y_u c_u-c_Q Y_u\right),\quad c_{QHd}=i\left(Y_d c_d-c_Q Y_d\right),\quad c_{LHe}=i\left(Y_e c_e-c_L Y_e\right)\,,
\end{align}
and also note that the axion gauge couplings are also shifted
\begin{align}
    &c_{\widetilde{G}}\rightarrow c_{\widetilde{G}}+\frac{1}{2}\text{Tr}\left(c_u+c_d-2c_Q\right) \\
    &c_{\widetilde{W}}\rightarrow c_{\widetilde{W}}-\frac{1}{2}\text{Tr}\left(3c_Q+c_L\right) \\
    &c_{\widetilde{B}}\rightarrow c_{\widetilde{B}}+\text{Tr}\left[3\left(y_u^2 c_u+y_d^2 c_d-2y_Q^2 c_Q\right)+\left(y_e^2 c_e-2y_L^2 c_L\right)\right]\,,
\end{align}
where $y_{Q, L, u, d, e}$ denote the hypercharges of the SM quarks and leptons.

Though Eq.~(\ref{eq:Laxionyukawa}) and Eq.~(\ref{eq:Laxionshift}) are equivalent, it is well-known that the non-derivative axion-fermion interaction terms do not form a non-redundant basis~\cite{Chala:2020wvs, Bauer:2020jbp, Bonilla:2021ufe}. This can be easily seen from the numbers of parameters of axion-fermion couplings, that Eq.~(\ref{eq:Laxionyukawa}) has total $3\times 18=54$ free parameters while Eq.~(\ref{eq:Laxionshift}) only has $5\times 9-4=41$ ones.

In this section, we are going to list higher dimension ($d\geq 5$) \gEFT~operators by imposing Adler's zero condition on a scalar singlet. Though the axion (or ALPs) is assumed to be a pseudoscalar, the CP-violating terms in the Lagrangian bury its CP property. Therefore one should note that without introducing further symmetries axion or ALP can not be distinguished from a general Goldstone. Therefore we use Goldstone and axion, \gEFT~and \aEFT~interchangeably. For the operators with more than $4$ fields, Adler's zero condition simply requires the scalar field comes with a derivative on it, which explicitly obeys shift symmetry. For the operators with $3$ or $4$ fields, the on-shell method usually can not generate derivative operators. Therefore, we have to check whether they obey Adler's zero condition or whether they can be reformulated into derivative forms by hand. We find that there are $5$ dangerous classes of operators, $dim$-5: $aF^2$ and $a\phi\psi^2$, $dim$-6: $aF\psi^2$ and $a\phi F^2$, and $dim$-7: $aF^3$, among which only the two $dim$-5 operator classes can be rewritten into derivative forms and contribute to the \aEFT.

\label{hyref:majoron}The Goldstone appears from the spontaneous breakdown of a hidden $U(1)$ lepton symmetry or $U(1)_{B-L}$ symmetry is also named as majoron which mediates lepton number violating process~\cite{Gelmini:1980re}. In the simplest realization, the majoron $J$ resides in a singlet complex scalar which carries $B-L$ charge 2. Besides the scalar, the simplest model also needs at least one right-handed neutrino $N_R$, which realizes the Type-I seesaw model~\cite{Minkowski:1977sc, Yanagida:1979as, Gell-Mann:1979vob, Mohapatra:1979ia}. The Lagrangian at renormalizable level reads
\begin{align}
    \mathcal{L}\supset&\overline{N_R} i\slashed{\partial}N_R+\partial_\mu\phi^\dagger\partial^\mu\phi-(y_R\overline{L}\widetilde{H}N_R+\frac{1}{2}\lambda\phi\overline{N_R^c} N_R+\text{h.c.}) \nonumber \\
    =&\overline{N_R} i\slashed{\partial}N_R-\frac{\lambda f/\sqrt{2}}{2}\left(\overline{N_R^c} N_R+\text{h.c.}\right)+\frac{1}{2}\partial_\mu J\partial^\mu J-(y_R\overline{L}\widetilde{H}N_R+\frac{1}{2}\frac{i\lambda}{\sqrt{2}}J\overline{N_R^c} N_R+\text{h.c.}) \nonumber \\
    &+\frac{1}{2}\partial_\mu\sigma\partial^\mu\sigma-\frac{1}{2}m_\sigma^2\sigma^2-\left(\frac{1}{2}\frac{\lambda}{\sqrt{2}}\sigma\overline{N_R^c} N_R+\text{h.c.}\right)\,,
\end{align}
where we ignore the potential of the scalar since it is irrelevant to our discussion. Once the scalar $\phi=(f+\sigma+iJ)/\sqrt{2}$ acquires non-vanishing VEV $f$ breaking the $B-L$ symmetry, the right-handed neutrino obtains a majorana mass $\lambda f$ leaving a massless goldstone, majoron $J$ and a heavy CP-even scalar $\sigma$ with a mass around $\mathcal{O}(f)$. Assuming this breaking scale is much higher than the typical SM scale, we could integrate out both the right-handed neutrino and CP-even scalar to obtain an EFT with only SM particles and majoron. For simplicity, we write down the EFT without the CP-even scalar contributions since 1) one can choose the potential of the scalar to set the mass of $\sigma$ much heavier and make the contributions from integrating $\sigma$ tiny, and 2) integrating $\sigma$ does not contribute to majoron interactions once ignoring scalar potential,
\begin{gather}
    \mathcal{L}\supset\left(c_{\nu\nu}\overline{L^c}\widetilde{H}^*\widetilde{H}^\dagger L+c_J J\overline{L^c}\widetilde{H}^*\widetilde{H}^\dagger L+\text{h.c.}\right)+c_{HL}\left[\left(H^\dagger i\overleftrightarrow{D}_\mu H\right)\left(\overline{L}\gamma^\mu L\right)-\left(H^\dagger i\overleftrightarrow{D}^I_\mu H\right)\left(\overline{L}\sigma^I\gamma^\mu L\right)\right]\,, \nonumber \\
    \text{w/ } c_{\nu\nu}=-\frac{y_R^2}{\sqrt{2}\lambda f},\qquad c_J=-i\frac{y_R^2}{\sqrt{2}\lambda f^2},\qquad c_{HL}=\frac{y_R^2}{2\lambda^2 f^2}\,, \label{eq:L_EFT_majoron}
\end{gather}
where the first term is just the $dim$-5 Weinberg operator~\cite{Weinberg:1979sa}, the last two terms are another two $dim$-6 operators in Warsaw basis~\cite{Grzadkowski:2010es}, and only the second term describes the interaction between the majoron and SM particles. Note that this operator is not shift invariant explicitly. As shown in the case of axion, we could use EOMs and IBP to rewrite it into a shift symmetry manifest form\footnote{Generally speaking, this procedure is non-trivial due to the dimensional parameter $\mu_H^2$ in the SM. And to the best of the author's knowledge, there is no reference in which explicitly shift-symmetric operators for majoron are given. Here we just schematically show that one can transform this operator into an operator given in our list. Dictionaries for the complete basis transformation like the one given for axion and the UV-EFT matching are left for the future publication.},
\begin{align}
    J\overline{L^c}\widetilde{H}^*\widetilde{H}^\dagger L\longrightarrow\frac{1}{\mu_H^2}{\epsilon^{ik}}{\epsilon^{jl}}({D^{\mu}}J){H_{k}}({D_{\mu}}{H_{l}})({{L}_{i}}{{L}_{j}})\,,
\end{align}
where $\mu_H^2$ is the mass parameter of the Higgs doublet in the SM Lagrangian.

The complete set of non-redundant operators for axion (Goldstone) and majoron can be read from Tab.~\ref{tab:ops4axion} in column $\textbf{Axion}$ and $\textbf{Majoron}$ respectively. For $dim$-5 operators of the axion, the operators are written into a form without derivative. Therefore, all these operators are mark as $\xcheckmark$ and should be understood via a basis transformation discussed above.
{ 
\begin{longtable}{cccccc}
\caption{\textcolor{black}{{Complete and independent operators for ALP (\textbf{Axion}), majoron (\textbf{Majoron}) classified in canonical dimension. Column \textbf{F} shows the flavor structure in Young tableau if repeated fields in operators lead to ambiguity. Complete and independent operators for real (\textbf{Real}) scalar are also presented for comparison. In order to obtain the non-redundant operators, a further basis transformation should be performed on the $dim$-5 operators with $\xcheckmark$ as described in the text.\label{tab:ops4axion}}}}\\
\hhline{======}
\multicolumn{6}{c}{\textcolor{blue}{{\textbf{\textit{dim}-5 operators}}}}\\
\hhline{======}
{\textbf{Class}} & {\textbf{Type}} & {\textbf{Real}} & {\textbf{F}} & {\textbf{Axion}} & {\textbf{Majoron}} \\
\hline\hline
\multirow{3}{*}{
} & $B_L^2 s$ & $s{{B_L}_{\mu\nu}}{{B_L}^{\mu\nu}}$ &  & $\xcheckmark$ &  \\ \cline{2-6}
\multirow{1}{*}{{$F_L^2\phi$}} & $W_L^2 s$ & $s{{{W_L}^{I}}_{\mu\nu}}{{{W_L}^{I}}^{\mu\nu}}$ &  & $\xcheckmark$ &  \\ \cline{2-6}
 & $G_L^2 s$ & $s{{{G_L}^{A}}_{\mu\nu}}{{{G_L}^{A}}^{\mu\nu}}$ &  & $\xcheckmark$ &  \\ \hline
\multirow{3}{*}{
} & $e_c H^\dagger L s$ & $s{{H^\dagger}^{i}}({{e_c}_{p}}{{L_{r}}_{i}})$ &  & $\xcheckmark$ &  \\ \cline{2-6}
\multirow{1}{*}{{$\psi^2 \phi^2$}} & {$d_c H^\dagger Q s$} & $s{{H^\dagger}^{i}}({{{d_c}_{p}}^{a}}{{Q_{r}}_{ai}})$ &  & $\xcheckmark$ &  \\
\hline
\end{longtable}
}
We further comment that it is better and much clearer to build the Goldstone EFT operators with the non-linear form of the Goldstone bosons instead of the Goldstone fields themselves due to the nature of spontaneous symmetry breaking\footnote{Though it is hard to see the advantages for axion case, the non-linear form helps to reduce the redundant operators if the Goldstones have flavor structure.}. And as you are going to see in Sec.~\ref{sec:operator4photon}, the non-linear form is also useful to build the EFT operators for dark photon. Therefore, here we also list the operators which contribute to $dim$-8 operators of Goldstone in non-linear form\footnote{For simplicity, we ignore the Wilson coefficients but just list the operators.},
\begin{align}
    \label{eq:LGoldstone}\mathcal{L}_{\text{GB}}\supset &u_\mu u^\mu+u_\mu u^\mu u_\nu u^\nu+{H_{i}}{{H^\dagger}^{i}}u_\mu u^\mu+{{H_{i}}{H_{j}}{{H^\dagger}^{i}}{{H^\dagger}^{j}}u_\mu u^\mu} \nonumber \\

    &+{{{B_L}_{\nu\lambda}}{{B_L}^{\lambda\nu}}u_\mu u^\mu}+{{{{W_L}^{I}}_{\nu\lambda}}{{{W_L}^{I}}^{\lambda\nu}}u_\mu u^\mu}+{{{{G_L}^{A}}_{\nu\lambda}}{{{G_L}^{A}}^{\lambda\nu}}u_\mu u^\mu} \nonumber \\
    
    &+{{{B_R}^{\lambda\mu}}{{{B_L}_{\lambda}}^{\nu}}u_\mu u_\nu}+{{{{W_R}^{I}}^{\lambda\mu}}{{{{W_L}^{I}}_{\lambda}}^{\nu}}u_\mu u_\nu}+{{{{G_R}^{A}}^{\lambda\mu}}{{{{G_L}^{A}}_{\lambda}}^{\nu}}u_\mu u_\nu} \nonumber \\

    &+{{\epsilon^{ij}}{H_{j}}({{Q_{p}}_{ai}}{{{u_c}_{r}}^{a}})u_\mu u^\mu}+{{{H^\dagger}^{i}}({{{d_c}_{p}}^{a}}{{Q_{r}}_{ai}})u_\mu u^\mu}+{{{H^\dagger}^{i}}({{e_c}_{p}}{{L_{r}}_{i}})u_\mu u^\mu} \nonumber \\
    
    &+(({D^{\mu}}{{Q_{p}}_{ai}}){\sigma^{\nu}}{{{Q^\dagger}_{r}}^{ai}})u_\mu u_\nu+(({D^{\mu}}{{{u_c}_{p}}^{a}}){\sigma^{\nu}}{{{u^\dagger_c}_{r}}_{a}})u_\mu u_\nu+(({D^{\mu}}{{{d_c}_{p}}^{a}}){\sigma^{\nu}}{{{d^\dagger_c}_{r}}_{a}})u_\mu u_\nu \nonumber \\

    &+(({D^{\mu}}{{e_c}_{p}}){\sigma^{\nu}}{{e^\dagger_c}_{r}})u_\mu u_\nu+(({D^{\mu}}{{L_{p}}_{i}}){\sigma^{\nu}}{{{L^\dagger}_{r}}^{i}})u_\mu u_\nu+{{H^\dagger}^{i}}({D^{\mu}}{D^{\nu}}{H_{i}})u_\mu u_\nu+({D_{\nu}}{H_{i}})({D^{\nu}}{{H^\dagger}^{i}})u_\mu u^\mu \nonumber \\

    &+{{H_{j}}{{H^\dagger}^{i}}({{Q_{p}}_{ai}}{\sigma_{\mu}}{{{Q^\dagger}_{r}}^{aj}})u^\mu}+{{H_{j}}{{H^\dagger}^{j}}({{Q_{p}}_{ai}}{\sigma_{\mu}}{{{Q^\dagger}_{r}}^{ai}})u^\mu}+{{H_{i}}{{H^\dagger}^{i}}({{{u_c}_{p}}^{a}}{\sigma_{\mu}}{{{u^\dagger_c}_{r}}_{a}})u^\mu} \nonumber \\

    &+{{H_{i}}{{H^\dagger}^{i}}({{{d_c}_{p}}^{a}}{\sigma_{\mu}}{{{d^\dagger_c}_{r}}_{a}})u^\mu}+{{H_{j}}{{H^\dagger}^{i}}({{L_{p}}_{i}}{\sigma_{\mu}}{{{L^\dagger}_{r}}^{j}})u^\mu}+{{H_{j}}{{H^\dagger}^{j}}({{L_{p}}_{i}}{\sigma_{\mu}}{{{L^\dagger}_{r}}^{i}})u^\mu}+{{H_{i}}{{H^\dagger}^{i}}({{e_c}_{p}}{\sigma_{\mu}}{{e^\dagger_c}_{r}})u^\mu} \nonumber\\

    &+{{{B_L}^{\mu\nu}}({{Q_{p}}_{ai}}{\sigma_{\mu}}{{{Q^\dagger}_{r}}^{ai}})u_\nu}+{{{B_L}^{\mu\nu}}({{{u_c}_{p}}^{a}}{\sigma_{\mu}}{{{u^\dagger_c}_{r}}_{a}})u_\nu}+{{{B_L}^{\mu\nu}}({{{d_c}_{p}}^{a}}{\sigma_{\mu}}{{{d^\dagger_c}_{r}}_{a}})u_\nu}+{{{B_L}^{\mu\nu}}({{L_{p}}_{i}}{\sigma_{\mu}}{{{L^\dagger}_{r}}^{i}})u_\nu} \nonumber \\

    &+{{{B_L}^{\mu\nu}}({{e_c}_{p}}{\sigma_{\mu}}{{e^\dagger_c}_{r}})u_\nu}+{{{\tau^{I}}_{j}^{i}}{{{W_L}^{I}}^{\mu\nu}}({{Q_{p}}_{ai}}{\sigma_{\mu}}{{{Q^\dagger}_{r}}^{aj}})u_\nu}+{{{\tau^{I}}_{j}^{i}}{{{W_L}^{I}}^{\mu\nu}}({{L_{p}}_{i}}{\sigma_{\mu}}{{{L^\dagger}_{r}}^{j}})u_\nu} \nonumber \\

    &+{{{\lambda^{A}}_{b}^{a}}{{{G_L}^{A}}^{\mu\nu}}({{Q_{p}}_{ai}}{\sigma_{\mu}}{{{Q^\dagger}_{r}}^{bi}})u_\nu}+{{{\lambda^{A}}_{a}^{b}}{{{G_L}^{A}}^{\mu\nu}}({{{u_c}_{p}}^{a}}{\sigma_{\mu}}{{{u^\dagger_c}_{r}}_{b}})u_\nu}+{{{\lambda^{A}}_{a}^{b}}{{{G_L}^{A}}^{\mu\nu}}({{{d_c}_{p}}^{a}}{\sigma_{\mu}}{{{d^\dagger_c}_{r}}_{b}})u_\nu} \nonumber \\

    &+{\mathcal{Y}[\tiny{\ytableaushort{pr}}]{\epsilon_{abc}}({{{d_c}_{p}}^{a}}{{L_{s}}_{i}})({{{d_c}_{r}}^{b}}{\sigma_{\mu}}{{{Q^\dagger}_{t}}^{ci}})u^\mu}+{\mathcal{Y}[\tiny{\ytableaushort{p,r}}]{\epsilon_{abc}}({{{d_c}_{p}}^{a}}{{L_{s}}_{i}})({{{d_c}_{r}}^{b}}{\sigma_{\mu}}{{{Q^\dagger}_{t}}^{ci}})u^\mu} \nonumber \\

    &+{\mathcal{Y}[\tiny{\ytableaushort{pr}}]{\epsilon^{ij}}({{{d_c}_{p}}^{a}}{{L_{r}}_{i}})({{L_{s}}_{j}}{\sigma_{\mu}}{{{u^\dagger_c}_{t}}_{a}})u^\mu}+{\mathcal{Y}[\tiny{\ytableaushort{p,r}}]{\epsilon^{ij}}({{{d_c}_{p}}^{a}}{{L_{r}}_{i}})({{L_{s}}_{j}}{\sigma_{\mu}}{{{u^\dagger_c}_{t}}_{a}})u^\mu}+{\epsilon_{abc}}({{{d_c}_{p}}^{a}}{{{d_c}_{r}}^{b}})({{{d_c}_{s}}^{c}}{\sigma_{\mu}}{{e^\dagger_c}_{t}})u^\mu \nonumber \\
    
    &+{H_{i}}{H_{j}}{{H^\dagger}^{i}}({D_{\mu}}{{H^\dagger}^{j}})u^\mu+{H_{i}}{{B_L}^{\mu\nu}}({D_{\mu}}{{H^\dagger}^{i}})u_\nu+{{\tau^{I}}_{j}^{i}}{H_{i}}{{{W_L}^{I}}^{\mu\nu}}({D_{\mu}}{{H^\dagger}^{j}})u_\nu  \nonumber \\

    &+({D_{\mu}}{{H^\dagger}^{i}})({{e_c}_{p}}{{L_{r}}_{i}})u^\mu+({D^{\mu}}{{H^\dagger}^{i}})({{e_c}_{p}}{\sigma_{\mu\nu}}{{L_{r}}_{i}})u^\nu+({D_{\mu}}{{H^\dagger}^{i}})({{{d_c}_{p}}^{a}}{{Q_{r}}_{ai}})u^\mu \nonumber \\
    
    &+({D^{\mu}}{{H^\dagger}^{i}})({{{d_c}_{p}}^{a}}{\sigma_{\mu\nu}}{{Q_{r}}_{ai}})u^\nu+{\epsilon^{ij}}({D_{\mu}}{H_{j}})({{Q_{p}}_{ai}}{{{u_c}_{r}}^{a}})u^\mu+{\epsilon^{ij}}({D^{\mu}}{H_{j}})({{Q_{p}}_{ai}}{\sigma_{\mu\nu}}{{{u_c}_{r}}^{a}})u^\nu \nonumber \\ 

    &+{\mathcal{Y}[\tiny{\ytableaushort{pr}}]{\epsilon^{ik}}{\epsilon^{jl}}{H_{k}}({D_{\mu}}{H_{l}})({{L_{p}}_{i}}{{L_{r}}_{j}})u^\mu}+{\mathcal{Y}[\tiny{\ytableaushort{p,r}}]{\epsilon^{ik}}{\epsilon^{jl}}{H_{k}}({D_{\mu}}{H_{l}})({{L_{p}}_{i}}{{L_{r}}_{j}})u^\mu} \nonumber \\

    &+{\mathcal{Y}[\tiny{\ytableaushort{pr}}]{\epsilon^{ik}}{\epsilon^{jl}}{H_{k}}({D^{\mu}}{H_{l}})({{L_{p}}_{i}}{\sigma_{\mu\nu}}{{L_{r}}_{j}})u^\nu}+{\mathcal{Y}[\tiny{\ytableaushort{p,r}}]{\epsilon^{ik}}{\epsilon^{jl}}{H_{k}}({D^{\mu}}{H_{l}})({{L_{p}}_{i}}{\sigma_{\mu\nu}}{{L_{r}}_{j}})u^\nu} \nonumber \\
    
    &+\text{h.c.}\,,
\end{align}
where $u_\mu\equiv iUD_\mu U^\dagger$. $U$ is the Goldstone matrix, which reduces to a single phase, $\exp(is/f)$, in our single Goldstone case with that $f$ is the breaking scale of the original Global symmetry~\footnote{Generally, we use $\mathbf{U}$ to represent a Goldstone matrix with arbitrary number of Goldstone fields. However, when it comes to the case with a single Goldstone field, we use $U$ instead to emphasize that it is now a phase and interchangeable with other fields.}. One can expand the exponential to arrive at the operators listed in the table at the leading order. 

\section{Dark Photon EFT Operators: Stueckelberg and Higgs Mechanism\label{sec:operator4photon}}
The singlet vector field is also called $U(1)$ dark photon~\cite{Okun:1982xi, Galison:1983pa, Holdom:1985ag}, which is commonly described by the Proca Lagrangian~\cite{Proca:1936fbw} with an addition kinetic mixing term to hypercharge
\begin{align}
    \mathcal{L}\supset-\frac{1}{4}A_{\mu\nu}A^{\mu\nu}+\frac{1}{2}m_X^2 A_\mu A^\mu-\epsilon B_{\mu\nu}A^{\mu\nu}\,, \label{eq:Proca}
\end{align}
where $A$ and $B$ are vector fields in gauge basis. Such a massive vector can directly account for part or whole of dark matter either via~\cite{Nelson:2011sf, Arias:2012az, Graham:2015rva}, or it can work as a portal if there are dark matter particles charged under this $U(1)$ symmetry~\cite{Knapen:2017xzo, Hambye:2019dwd, Fabbrichesi:2020wbt, Alexander:2016aln, Battaglieri:2017aum}.

However, it is well-known that the Proca Lagrangian is not gauge invariant and needs to be UV completed. There are two methods to achieve this, 1) Higgs mechanism~\cite{Higgs:1964pj, Englert:1964et} and 2) Stueckelberg mechanism~\cite{Stueckelberg:1938hvi, Ruegg:2003ps}. Since our following discussion will be based on the Stueckelberg mechanism, we will first review the Stueckelberg mechanism, and then comment on the differences and similarities between Stueckelberg and Higgs mechanism.

The original Stueckelberg Lagrangian for massive vector $X_\mu$ can be written as
\begin{align}
    \mathcal{L}_\text{Stueck}=&-\frac{1}{4}X_{\mu\nu}X^{\mu\nu}+\frac{1}{2}m_X^2(X_\mu-\frac{1}{m_X}\partial_\mu\pi)(X^\mu-\frac{1}{m_X}\partial^\mu\pi) \nonumber \\
    &-\frac{1}{2}(\partial^\mu X_\mu+m_X\pi)(\partial^\nu X_\nu+m_X\pi)\,, \label{eq:L_Stueck}
\end{align}
where $\pi$ is the Stueckelberg field. Under $U(1)$ gauge transformation,
the fields transform as
\begin{align}
    X_\mu&\longrightarrow X_\mu+\partial_\mu\alpha(x)\,, \\
    \pi&\longrightarrow\pi+m_X\alpha(x)\,,
\end{align}
where $\alpha(x)$ is the gauge parameter. The first line of Eq.~(\ref{eq:L_Stueck}) is invariant under the above transformation, while the second line should be identified as a gauge fixing term which removes the mixing terms of the form $X_\mu\partial_\mu B$ after expansion and can be generalized to 't Hooft gauge-fixing term
\begin{align}
    \mathcal{L}'_\text{Stueck}=&-\frac{1}{4}X_{\mu\nu}X^{\mu\nu}+\frac{1}{2}m_X^2(X_\mu-\frac{1}{m_X}\partial_\mu\pi)(X^\mu-\frac{1}{m_X}\partial^\mu\pi) \nonumber \\
    &-\frac{1}{2\xi}(\partial^\mu X_\mu+\xi m_X\pi)(\partial^\nu X_\nu+\xi m_X\pi)\,.
\end{align}
From this, one can see the Stueckelberg Lagrangian is obtained from the choice of Feynman gauge $\xi=1$, in which the transverse polarizations $X_\mu$ and longitudinal polarization $\pi$ (or $\partial_\mu\pi$) are completely separate~\footnote{To be more rigorous, ghost fields has to be introduced in the general gauge-fixing and quantization procedure. In Feynman gauge, the ghosts remove the longitudinal and timelike polarizations and make gauge field $X_\mu$ pure transverse. However, the ghosts decouple in Abelian gauge theories. Hence  we omit them from the discussion.}. Provided that 't Hooft-like gauge functional $\mathcal{G}=\partial^\mu X_\mu+\xi m_X\pi$ vanishes, one recovers the Proca theory Eq.~(\ref{eq:Proca}) after transformation $X_\mu\rightarrow A_\mu+\partial_\mu\pi/m_X$. In such case, the gauge field $A_\mu$ absorbs the Stueckelberg field and becomes a field with three physical polarizations, which is also equivalent to the unitary gauge ($\xi=\infty$).

By defining $U\equiv e^{i\frac{\pi}{m_X}}$, we have
\begin{align}
    D_\mu U\equiv\left(\partial_\mu-iX_\mu\right)U=iU\left(\frac{1}{m_X}\partial_\mu\pi-X_\mu\right)\,.
\end{align}
The generalized Stueckelberg Lagrangian can be further rewritten into the form of
\begin{align}
    \mathcal{L}'_\text{Stueck}=&-\frac{1}{4}X_{\mu\nu}X^{\mu\nu}+\frac{m_X^2}{2}(D_\mu U)^\dagger D^\mu U-\frac{1}{2\xi}(\partial^\mu X_\mu-i\xi m_X^2\ln U)^2\,,
\end{align}
which has the exact same form of the leading order chiral Lagrangian for $U(1)$ gauge field~\cite{Zhang:2008xq}. One can identify the matrix field $U$ as the Goldstone matrix generated from $U(1)_L\times U(1)_R\rightarrow U(1)_V$. Under such symmetry, $U$ field transforms as
\begin{align}
    U\longrightarrow RUL^\dagger=RL^\dagger U\,,
\end{align}
where $L$ and $R$ are the group elements of $U(1)_L$ and $U(1)_R$ respectively. The original gauged $U(1)$ symmetry can now be identified as $U(1)_V$. Choosing unitary gauge $\xi=\infty$ ($U=1$), the second term just becomes the vector boson massive term $m_X^2 X_\mu X^\mu/2$. Therefore the Stueckelberg Lagrangian involves two parts, 1) gauge invariant terms and 2) gauge fixing term. Noticing this fact, one can obtain the EFT via writing down the gauge invariant terms built with vector field and Goldstone (Stueckelberg) field, while gauging fixing term can be ignored and added trivially as one does in electroweak chiral Lagrangian~\cite{Longhitano:1980tm, Longhitano:1980iz, Feruglio:1992wf, Herrero:1993nc}.

In Higgs mechanism, a $U(1)$ charged complex scalar, $S$, with a Mexican hat potential is introduced
\begin{align}
    \mathcal{L}_{\text{Dark }\gamma+\text{Dark Higgs}}=&-\frac{1}{4}X_{\mu\nu}X^{\mu\nu}+D_\mu S^\dagger D^\mu S+\mu^2 S^\dagger S-\lambda\left(S^\dagger S\right)^2\,,
\end{align}
where $S$ can develop a non-vanishing VEV $\braket{S}\equiv v_S/\sqrt{2}=\mu/\sqrt{2\lambda}$. To compared to the Stueckelberg mechanism, we write field $S$ in polar coordinates
\begin{align}
    S=\frac{1}{\sqrt{2}}\left(v_S+\sigma\right)e^{-i\frac{\pi}{v_S}}\,,
\end{align}
and the Lagrangian is now written as
\begin{equation}
\begin{split}
    \mathcal{L}_{\text{Dark }\gamma+\text{Dark Higgs}}=&-\frac{1}{4}X_{\mu\nu}X^{\mu\nu}+\frac{1}{2}g^2 v_S^2\left(X_\mu-\frac{1}{gv_S}\partial_\mu\pi\right)\left(X^\mu-\frac{1}{gv_S}\partial^\mu\pi\right) \\
    &+\frac{1}{2}\partial_\mu\sigma\partial^\mu\sigma-\frac{1}{2}2\lambda v_S^2\sigma^2-\lambda v\sigma^3-\frac{1}{4}\lambda\sigma^4 \\
    &+\left(\frac{\sigma}{v}+\frac{\sigma^2}{2v^2}\right)\left(gvX_\mu-\partial_\mu\pi\right)\left(gvX^\mu-\partial^\mu\pi\right)\,,
\end{split}
\end{equation}
where the first line is exactly the same as the gauge invariant part in the Stueckelberg Lagrangian once we identify the dark photon mass $m_X$ as $gv_S$. While the second line includes a real scalar (dark Higgs) with mass $m_\sigma=2\lambda v_S^2$, trilinear and quartic self interactions, the last line describes the interactions between dark Higgs $\sigma$ and dark photon $X_\mu$ and/or Goldstone boson $\pi$. We can recognize the Stueckelberg mechanism as a special case of the Higgs mechanism where $\sigma$ is decoupled through the limit $\lambda\rightarrow\infty$ with $v_S$ fixed.

Integrating out the Higgs field $\sigma$ at the tree level, we arrive at an effective theory with only vector and Goldstone fields
\begin{align}
    \mathcal{L}_{\text{Dark }\gamma+\text{Dark Higgs}-\sigma}^\text{EFT}=&-\frac{1}{4}X_{\mu\nu}X^{\mu\nu}+\frac{1}{2}g^2 v_S^2\left(X_\mu-\frac{1}{gv_S}\partial_\mu\pi\right)\left(X^\mu-\frac{1}{gv_S}\partial^\mu\pi\right) \nonumber \\
    &+\frac{g^4 v_S^2}{2m_\sigma^2}\left[\left(X_\mu-\frac{1}{gv_S}\partial_\mu\pi\right)\left(X^\mu-\frac{1}{gv_S}\partial^\mu\pi\right)\right]^2+\cdots\,,
\end{align}
where we only keep the leading term in order of $\mathcal{O}(1/m_\sigma^2)$. Note that there is a new $dim$-4 term $\left(X_\mu X^\mu\right)^2$ with coupling strength $g^4 v_S^2/2m_\sigma^2=g^4/4\lambda$, which is not present in the standard Stueckelberg Lagrangian. This can be understood that the Stueckelberg Lagrangian is equivalent to the infinite mass limit (equivalently infinite $\lambda$ limit) of Higgs mechanism, while a non-natural mass hierarchy can exist between dark photon and dark Higgs (especially in a strongly interacting model, large but finite $\lambda$), which leads to a quartic term in a EFT with only dark photon. Such a term has also been discussed in Ref.~\cite{Reece:2018zvv, Kribs:2022gri} with their own purposes. However, here we emphasize that such a term must be included in a complete basis from an EFT point of view. And our gauge invariant method indeed generates it at the $dim$-4 level.

Based on the above discussion, we can obtain a gauge invariant massive dark photon theory via building blocks, tensor field $X_{\mu\nu}$ (and its dual $\widetilde{X}_{\mu\nu}$) and a Goldstone matrix $U$. As discussed in Sec.~\ref{sec:opsbbs&YTT} the EFT operators involving Goldstones can be systematically built via contact on-shell amplitude obtained from Young tensor method by imposing the Adler zero condition. We apply same technique here to write down dark photon EFT (\XEFT) operator basis listed in Tab.~\ref{tab:ops4DP}. However, there are two main differences commented below,
\begin{itemize}
    \item Building Block, $U$: since $U$ is an exact Goldstone, it always comes with a covariant derivative. However instead to use $D_\mu U$ discussed in this section as our building block, we are going to use $u_\mu\equiv i UD_\mu U$, which is also a gauge invariant term due to the fact that $U(1)$ transformation is only a multiplication by phase factor.

    \item Power Counting: unlike the canonical dimension used in the SMEFT~\cite{Weinberg:1979sa, Buchmuller:1985jz, Grzadkowski:2010es, Lehman:2014jma, Liao:2016hru, Li:2020gnx, Murphy:2020rsh, Li:2020xlh, Liao:2020jmn}, we use the chiral dimension counting like in chiral perturbation theory and HEFT~\cite{Weinberg:1978kz, Gasser:1983yg, Manohar:1983md, Gasser:1984gg, Hirn:2005fr, Buchalla:2012qq, Buchalla:2013eza, Pich:2015kwa, Gavela:2016bzc, Buchalla:2016sop, Pich:2016lew, Krause:2018cwe}. We identify $u_\mu$ having dimension $1$ (chiral dimension), while all other building blocks (SM fields, $X_{\mu\nu}$ and $\widetilde{X}_{\mu\nu}$) use their standard canonical dimension. For example, the operator ${H_{i}}{{H^\dagger}^{i}}u_\mu u^\mu$ has two scalar fields $H$ and $H^\dagger$, and two $u_\mu$, thus we say this operator is a $dim$-4 operator. In unitary gauge, $u_\mu$ reduces to $-iX_\mu$, which has canonical dimension $1$. Therefore such power counting just counts the canonical dimension of the operators in unitary gauge. Our example operator is equivalent to ${H_{i}}{{H^\dagger}^{i}}X_\mu X^\mu$, which has also been considered in Ref.~\cite{Criado:2021trs, Kribs:2022gri}.
\end{itemize}

\setlength\LTleft{-0.45in}
{ \small 
\begin{longtable}{@{}C{0.09\textwidth}C{0.12\textwidth}C{0.36\textwidth}C{0.365\textwidth}C{0.1\textwidth}@{}}
\caption{\textcolor{black}{{Complete and independent operators for dark photon written in Stueckelberg mechanism with Goldstone (\textbf{Stueckelberg}), unitary gauge with massive vector field (\textbf{Unitary}) classified in canonical dimension.~\label{tab:ops4DP}}}}\\
\hhline{=====}
\multicolumn{5}{c}{\textcolor{blue}{{\textbf{\textit{dim}-4 operators}}}}\\
\hhline{=====}
{\textbf{Class}} & {\textbf{Type}} & {\textbf{Stueckelberg}} & {\textbf{Unitary}} & {\textbf{F}}  \\
\hline\hline
$F_L^2$ & $X_L B_L$ & ${B_L}_{\mu\nu}{X_L}^{\mu\nu}$ & $\checkmark$ &  \\ \hline

\multirow{5}{*}{
} & $u e_c e_c^\dagger$ & $({{e_c}_{p}}{\sigma^{\mu}}{{e^\dagger_c}_{r}})u_\mu$ & $({{e_c}_{p}}{\sigma^{\mu}}{{e^\dagger_c}_{r}})X_\mu$ &  \\ \cline{2-5}
 & $u LL^\dagger$ & $({{L_{p}}_{i}}{\sigma^{\mu}}{{{L^\dagger}_{r}}^{i}})u_\mu$ & $({{L_{p}}_{i}}{\sigma^{\mu}}{{{L^\dagger}_{r}}^{i}})X_\mu$ &  \\ \cline{2-5}
\multirow{1}{*}{$u\psi\bar{\psi}$} & $u QQ^\dagger$ & $({{Q_{p}}_{ai}}{\sigma^{\mu}}{{{Q^\dagger}_{r}}^{ai}})u_\mu$ & $({{Q_{p}}_{ai}}{\sigma^{\mu}}{{{Q^\dagger}_{r}}^{ai}})X_\mu$ &  \\ \cline{2-5}
 & $u d_c d_c^\dagger$ & $({{{d_c}_{p}}^{a}}{\sigma^{\mu}}{{{d^\dagger_c}_{r}}_{a}})u_\mu$ & $({{{d_c}_{p}}^{a}}{\sigma^{\mu}}{{{d^\dagger_c}_{r}}_{a}})X_\mu$ &  \\
\hline

\end{longtable}
}

\section{Conclusion \label{sec:conclu}}
In this paper, we present the independent and complete operators of a Goldstone singlet and a light dark photon extension of the SMEFT up to dimension-8 for the first time, by means of the on-shell amplitude method with Young tensor technique. The operators involving the Goldstone bosons (\gEFT) can be reduced to the subspace of the basis for scalar extensions satisfying Adler's zero condition in the soft momentum limit. The Goldstone can be identified as the Stueckelberg field and works as the longitudinal mode in the SMEFT extension with an extra gauged $U(1)$ symmetry. Combining our Goldstone operator basis and a massless $U(1)$ gauge basis, the basis of dark photon extension (\XEFT) can be built in a gauge invariant way~\footnote{One could also build the basis of \XEFT~simply introducing a massive vector field satisfying only Lorentz symmetry and without considering the extra $U(1)$ symmetry. However, it is more complicated to get rid of the redundancy in such way. Our Unitary basis gives the equivalent operators and also reveals the underlying symmetry explicitly.}.

With the enumerated operators, we can also get the total numbers of the independent operators in the two EFTs. Following the original notation in Ref.~\cite{Li:2020gnx} we refer to a type of operators whose conjugate are of a different type as a “complex” type, and a self-conjugate type as a “real” type. Therefore each operator in the ``complex'' types should be counted twice in the sense of Hermitian degrees of freedom, while those in the “real” types are only counted as one degree of freedom. For the operators with repeated fields, one should include the contributions from the permutation symmetry of their flavor indices which can be read from  Column $\textbf{F}$. In the \gEFT, we obtain that there are 6 (44) $dim$-5 operators, 1 (1) $dim$-6 operator, 44 (356) $dim$-7 operators and 32 (520) $dim$-8 operators with one (three) generation of fermions with power counting based on canonical dimensions. In the \XEFT, we use chiral power counting by identifying $u_\mu$ with dimension 1 while keeping other building blocks having their standard canonical dimension. We find that there are 9 (49), 0 (0), 108 (676), 10 (426) and 1904 (40783) operators at the dimension 4, 5, 6, 7, 8 involving one (three) generation of fermion flavors. We also count these operators via the Hilbert series method, which gives consistent results.

Given the fact that a (pseudo-)Nambu-Goldstone is naturally light, this Goldstone extension (and its variations) benefits various phenomenological studies involving axion, ALPs or majoron, which might be underestimated in the SMEFT analyses. On the one hand, the complete bases are necessarily needed during matching between the UV physics and the effective theories; on the other hand, the experimental signals coming from the higher dimensional operators would be comparable to the ones from well-studied leading operators, especially in UV models where the leading operators are suppressed either by some symmetries or by accidental cancellations. Though Goldstone and dark photon provide us naturally light particles beyond the SM particles, accidental light scalars or fermions can also exist and we present complete sets of such extensions of the SMEFT in a companion paper~\cite{Song:2023jqm}.
  

	
\section*{Acknowledgments} 
We thank Ming-Lei Xiao, Hao-Lin Li and Yu-Hui Zheng for helpful discussions. This work is supported by the National Science Foundation of China under Grants No. 12022514, No. 11875003 and No. 12047503, and National Key Research and Development Program of China Grant No. 2020YFC2201501, No. 2021YFA0718304, and CAS Project for Young Scientists in Basic Research YSBR-006, the Key Research Program of the CAS Grant No. XDPB15. 



%


\bibliographystyle{JHEP}
\bibliography{ref}

\end{document}